\documentclass[useAMS,usenatbib]{mn2e}
\usepackage{graphicx}

\title[An H$\alpha$-selected sample of cataclysmic variables -- II.]{An H$\alpha$-selected sample of cataclysmic variables -- II. Implications for CV evolution}
\author[M.L. Pretorius and C. Knigge]{Magaretha L. Pretorius\thanks{E-mail: mlp@astro.soton.ac.uk (MLP); christian@astro.soton. ac.uk (CK)} and Christian Knigge\footnotemark[1] \\
School of Physics and Astronomy, University of Southampton, Highfield, Southampton SO17 1BJ, United Kingdom\\
}
\begin{document}

%\date{Accepted 1988 December 15. Received 1988 December 14; in original form 1988 October 11}

\pagerange{\pageref{firstpage}--\pageref{lastpage}} \pubyear{}

\maketitle

\label{firstpage}

\begin{abstract}
We use an independent new sample of cataclysmic variables (CVs), constructed by selecting objects for H$\alpha$ emission, to constrain the properties of the intrinsic CV population.  This sample is restricted to systems that are likely to be non-magnetic and unevolved; it consists of 17 CVs, of which at least 10 have orbital periods above 3~h.  We find that even very generous allowance for selection effects is not sufficient to reconcile the large ratio of short- to long-period CVs predicted by standard CV evolution theory with the observed sample, possibly implying that short-period systems evolve faster than predicted by the disrupted magnetic braking model.  This would require that an angular momentum loss mechanism, besides gravitational radiation, acts on CVs with orbital periods below the period gap.  To bring the model into agreement with observations, the rate of angular momentum loss below the period gap must be increased by a factor of at least 3, unless the model also over estimates the angular momentum loss rate of long-period CVs.  
\end{abstract}


\section{Introduction}
Despite its wide relevance, the evolution of cataclysmic variables (CVs) is still not very well understood.  Much of the problem stems from the fact that strong selection effects act on observed samples of CVs, making it difficult to constrain theory observationally.  

The angular momentum loss rate ($-\dot{J}$) is the crucial ingredient of CV evolution theory.  Angular momentum loss leads to mass transfer from the secondary to the white dwarf.  The relation between the mass transfer rate ($\dot{M}$) and $\dot{J}$ depends on the structure of the secondary star, and the reaction of the secondary to mass loss determines the orbital period ($P_{orb}$) evolution of the system.  

Population synthesis methods combine the changing $\dot{M}$ and $P_{orb}$ with a model of the birthrate of CVs to predict the present-day distribution of CVs as a function of $M_1$, $M_2$ (the mass of the white dwarf and secondary, respectively), $\dot{M}$, and $P_{orb}$.  Only one of these parameters, $P_{orb}$, can practically be measured for a large number of CVs (but see \citealt{Patterson98} for an indirect method to measure mass ratios).  The orbital period distribution of known CVs is therefore one of the few observational properties of the CV population that can be used to constrain theory.  The other obvious constraint is the observed CV space density.  

In comparing both the size and period distribution of the known CV population to theoretical predictions, observational bias must be accounted for.  This requires that observed CV samples be restricted to objects selected in well-defined (and preferably homogeneous) ways.

A long-standing problem is that theory predicts a large population of short-period CVs, consisting mostly of period bouncers (systems that have evolved beyond the period minimum at about 76~min), which is not observed.  The models of \cite{Kolb93} and \cite{HowellRappaportPolitano97} both predict that only $\simeq 1$\% of CVs are long-period systems, while roughly 70\% are period bouncers.  Existing observations have already been used to argue that the short-period CV population cannot be as large as predicted (e.g. \citealt{Patterson98}; \citealt{PretoriusKniggeKolb07}).  We will do the same here, using a new and independent CV sample.  

The CV samples that have been available to date are heavily biased against the intrinsically faint, short-period CVs, mainly because most have bright flux limits.  The new sample considered here is also limited to apparently bright systems.  However, almost all surveys incorporate a second selection cut (most commonly a blue cut) that also discriminates against the discovery of short-period CVs.  The CV sample that we have constructed differs from most existing samples in that the only selection criterion (other than a flux limit) is based on line emission.

The spectra of the majority of CVs show Balmer emission lines, originating mainly in the accretion flow.  The luminosity of CVs is anti-correlated to the equivalent widths (EWs) of their emission lines (\citealt{Patterson84}; \citealt{WithamKniggeGansicke06}; we will take EWs of emission lines as positive throughout).  The explanation for this anti-correlation is that intrinsically faint CVs are low-$\dot{M}$ systems with low density discs, in which recombination line cooling is very efficient.  Therefore, in contrast to other selection criteria (such as blue optical colours and variability), an emission line EW-based selection cut should favour the discovery of intrinsically faint, short-period systems.

We have used the AAO/UKST SuperCOSMOS H$\alpha$ Survey (SHS) to define a homogeneous sample of CVs, selected on the basis of H$\alpha$ emission.  Observations of the new CVs were presented in an earlier paper (\citealt{HalphaI}; hereafter Paper I).  Here we describe the construction and completeness of the sample, examine the observational biases affecting it, and compare it to the predictions of theory.

\section{Identifying CVs in the SHS}
The SHS is a photographic survey carried out with the UK Schmidt Telescope \citep{Parker05}.  Plates were scanned by the SuperCOSMOS digitizing machine, and the data are publicly available.  The survey covered a total of $\sim3\,700\,\mathrm{sq.deg.}$ (233 separate $4^\circ \times 4^\circ$ fields) at low Galactic latitude in the southern hemisphere ($|b| \la +10^\circ$ and $\delta < +2^\circ$), in $R$ and narrow-band H$\alpha$.  The limiting magnitude is $R\simeq 20.5$, and the angular resolution of the images is $\simeq 1''$.  

The digitized plates are processed to extract sources and provide Image Analysis Mode (IAM) data for the detected objects.  The analysis assigns, amongst other parameters, classification, quality, and blend flags to each source.  This makes it possible to obtain catalogues containing only well-isolated point sources with relatively good photometry.  

Older UKST broad-band surveys, including an $I$-band survey of the whole southern sky, have also been scanned by SuperCOSMOS (\citealt{supercosmosI}; \citealt{supercosmosII}; \citealt{supercosmosIII}).  The photometric accuracy of these surveys is discussed in \cite{supercosmosII}.  The SHS $R$ and H$\alpha$ data are expected to be of similar quality, since the survey was digitized and processed in the same way.  The photometry is calibrated to remove systematic errors in colour as a function of magnitude and position on a given plate; this is described in \cite{supercosmosII} and \cite{Parker05}.  The result is that errors in colours are typically smaller than would be expected from the errors in magnitudes.  By matching the UKST catalogues to CCD data from the INT/WFC photometric H$\alpha$ survey of the northern galactic plane (IPHAS; see \citealt{Drew05}) in regions near $\delta = 0^\circ$ where the two surveys overlap, we made a rough estimate of errors in the SHS colours.  Typical errors for stars brighter than $R=17$ are  $\simeq 0.07$ in $R-\mathrm{H}\alpha$ and $\simeq 0.14$ in $R-I$, although these errors vary from field to field.

Given the limited quality of photographic photometry, we do not expect any selection procedure to identify emission line objects in this survey with complete reliability.  Poor photometry is also not the only reason for the selection to fail.  Because H$\alpha$ and corresponding $R$-band plates are in many cases taken months or even years apart, and all $I$-band images were obtained decades earlier, variable stars can erroneously be included in the sample.  Also, since M stars have strong molecular absorption bands in both the $R$ and $I$ pass-bands, they are another class of objects that can be selected, even if no emission lines appear in their spectra.  Multi-object spectroscopy has been used for the follow-up of objects from this survey (e.g. \citealt{Hopewell05}), and has yielded low hit-rates for identifying emission line objects.  We therefore chose selection criteria that are quite conservative, and that target only objects with large $R-\mathrm{H}\alpha$ excess.

\subsection{The selection procedure}

\begin{figure}
 \includegraphics[width=84mm]{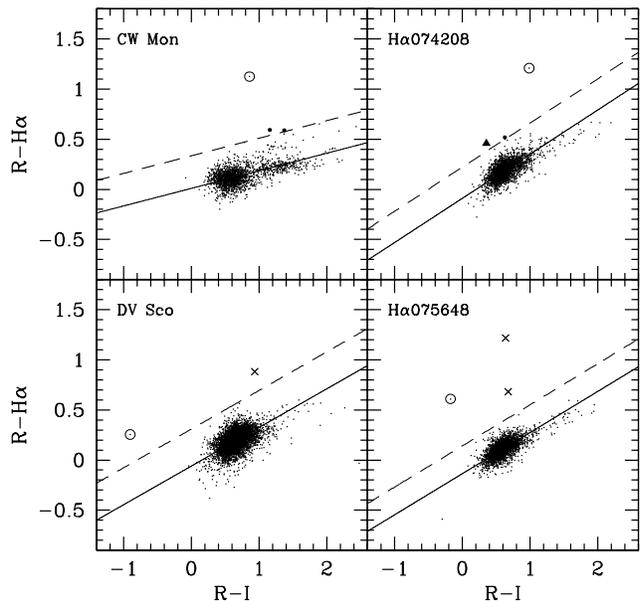}
 \caption{Colour-colour diagrams of $1^\circ \times 1^\circ$ sections of four of the SHS fields in which CVs were identified.  The solid lines are least-squares fits to all data points, while the dashed lines are vertically displaced by $4 \times$ the rms scatter in $R-\mathrm{H}\alpha$ around the best fits.  Only sources above the dashed lines are considered for selection.  Each panel is labeled with the name of the CV selected from the plot, and the colours of the CVs are plotted as circled points.  The other points above the cut-off are plate defects (crosses), objects that were not included in the sample, despite satisfying the cut-off (filled circles), and an object that was observed but has no H$\alpha$ emission (triangle).
}
 \label{fig:selection_cc}
\end{figure}

Our selection is aimed at bright objects that are clear outliers in the $R-\mathrm{H}\alpha$ vs $R-I$ colour-colour plane.  The selection procedure is similar to  that of \cite{WithamKniggeGansicke06}, but simpler (we use only a single magnitude bin, and do not attempt to isolate an unreddened main sequence, since this sequence is typically not separately discernible in the SHS photometry).

We considered 175 out of the 233 fields (i.e. $2\,800\,\mathrm{sq.deg.}$ in total).  These fields were chosen to have relatively good photometry (colour-colour diagrams of $2^\circ \times 2^\circ$ sections of fields were inspected, and in cases where one or more of these sections had a very poorly defined stellar locus, the whole field was discarded); fields observed in the multi-object spectroscopy program of which some results are presented by \cite{Hopewell05} were also avoided.

We extracted catalogues covering $4^\circ \times 4^\circ$ around every of the 175 field centres, and restricted these catalogues to unblended point sources brighter than $R=17.0$, and with good quality photometry in $R$, H$\alpha$, and $I$\footnote{To be exact, we considered objects that satisfied all of the following criteria ($R$ and H$\alpha$ magnitudes are here denoted SR and R\_Ha, respectively).  SR\,$<$\,17, R\_Ha\,$<$\,99, I\,$<$\,99, QUAL\_SR\,$<$\,128, QUAL\_Ha\,$<$\,128, QUAL\_I\,$<$\,128, BLEND\_SR\,$=$\,0, BLEND\_Ha\,$=$\,0, BLEND\_I\,$=$\,0, and CLASS\,$=$\,2.  The meaning of the quality flags (e.g. QUAL\_SR) is described by \cite{supercosmosII}; the flag CLASS\,$=$\,2 is assigned to objects classified as stellar.}.

Since the position of the stellar locus in the $R-\mathrm{H}\alpha$ vs $R-I$ plane changes, even across a single $4^\circ \times 4^\circ$ field (mainly because the effective photometric calibration is not constant), each field was divided into 16 $1^\circ \times 1^\circ$ sub-fields, and these 1~sq.deg. sub-fields were used individually in selecting H$\alpha$ excess sources.  

Fig.~\ref{fig:selection_cc} shows the $R-\mathrm{H}\alpha$ excess as a function of $R-I$ colour of objects in four such sub-fields.   The solid lines are linear fits to the colours of all objects; note that the stellar locus has a positive slope in all these plots.  The necessary, but not sufficient, criterion for selection is that targets have $R-\mathrm{H}\alpha>4 \times$ the rms scatter in $R-\mathrm{H}\alpha$ around the best linear fit to the stellar locus.  The dashed lines are vertically displaced by this amount from the best fits, and are the cut-off criteria for inclusion in the sample.  

In fields where not all objects above the cut-off are cleanly separated from the stellar locus, some objects satisfying the cut-off are not selected.  Examples of such objects are plotted as filled circles in Fig.~\ref{fig:selection_cc}.  This amounts to subjectively excluding some potential targets, based on inspection of the colour-colour diagrams (or, equivalently, using a stricter $R-\mathrm{H}\alpha$ selection cut in some fields).  

The colour-colour diagrams in Fig.~\ref{fig:selection_cc} are for fields from which two new, and two previously known CV were selected.  It illustrates that objects with fairly small $R-\mathrm{H}\alpha$ can be selected if they are blue in $R-I$ (e.g. DV Sco), and also that, because of the non-zero slope of the stellar locus, a selection algorithm based only on $R-\mathrm{H}\alpha$ will not successfully differentiate between true outliers and objects with red $R-I$.

A selection based on abnormal colours, not surprisingly, identifies a large number of objects with spurious photometry in one or more wave-band.  Therefore, we inspected the H$\alpha$-, $R$-, and $I$-band images of all potential targets to exclude objects that are outliers because of, e.g., plate defects (some of the various kinds of plate defects that affect the data are illustrated in \citealt{supercosmosI} and \citealt{Parker05}).  Outliers that were excluded from the target list for this reason are plotted as crosses in Fig.~\ref{fig:selection_cc}.  Next, we also excluded blends, objects from images that appear to have been taken in particularly poor seeing, and a few objects in areas with very high H$\alpha$ background.  Less than 30\% of potential targets survive these quality cuts.

With the exception of objects that are known only as X-ray sources, the targets that could be matched to known objects in the SIMBAD Astronomical Database were also excluded.  The known objects included planetary nebulae, T Tauri stars, CVs, and several other types of variable stars.

The selection produced 507 targets, and we obtained identification spectra for 460 of them (no additional criteria were applied to choose these 460 objects; we simply observed as many targets as time allowed for).  172 of the observed targets really are H$\alpha$ emission line objects (a hit-rate of only 37\%), and 14 (that is 3\%) are CVs.  Of the 288 targets with no H$\alpha$ emission, 10 are M stars (we also found many M stars with H$\alpha$ emission).  Some combination of poor survey photometry and variability (probably in most cases continuum, rather than emission line variability) led to the selection of the remaining 278 targets found to have no H$\alpha$ emission.  

\begin{figure}
 \includegraphics[width=84mm]{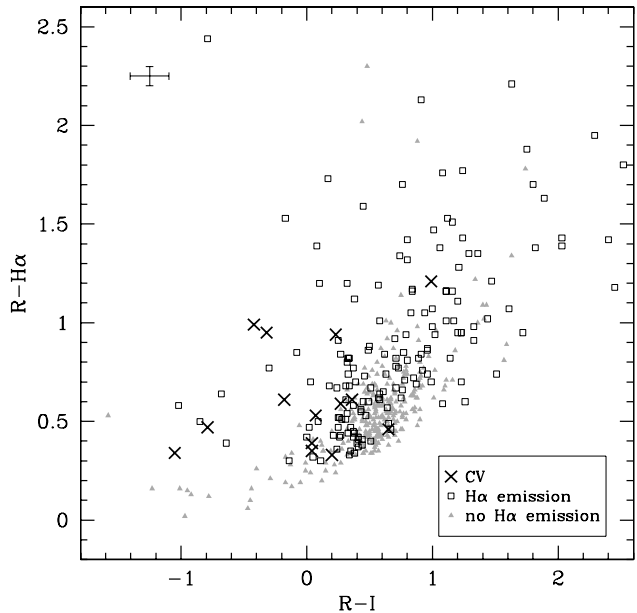}
 \caption{SHS colours of all objects for which we obtained identification spectra.  An estimate of typical 1$\sigma$ errors in these colours are shown in the top left hand corner.  Several objects lie outside the $R-I$ and $R-\mathrm{H}\alpha$ ranges plotted.  Note that the relative positions of objects in this plot can be misleading, since the position and slope of the stellar locus vary.
}
 \label{fig:allcolcol}
\end{figure}

Fig.~\ref{fig:allcolcol} shows $R-\mathrm{H}\alpha$ excess as a function of $R-I$ colour for all the objects we observed.  The colours of CVs are plotted as crosses; other emission line objects are shown as open square points, and objects without H$\alpha$ emission are plotted as grey triangles.  Note that some of the objects without H$\alpha$ emission have very large $R-\mathrm{H}\alpha$, and that the CVs are systematically bluer in $R-I$ than the overall target sample.

We also investigated other optical colours, as well as near-IR colours of the targets.  While it is possible to separate the majority of late type emission line stars from CVs using e.g. Two Micron All Sky Survey (2MASS; \citealt{2mass}) photometry, Be stars and other early type emission line stars, as well as the objects without any line emission (our most important contaminant), occupy the same regions as most CVs in all the colour-colour and colour-magnitude planes.  It will therefore not be easy to construct a much more efficient selection procedure to find CVs in this survey.  

In Paper~I we included two CVs fainter than $R=17$.  These were discovered during an observing run for which we experimented with selecting objects in a fainter magnitude bin.  We will here consider only CVs from the $R<17$ selection.

\subsection{Recovery of previously known CVs}
The fraction of known CVs recovered by our selection gives an indication of the completeness of the CV sample.  There are 42 previously known CVs with $R<17$ in the survey area covered by our follow-up, and only 7 of these were selected as targets.  The majority of the CVs that were not selected (30 systems) were excluded by the cuts in the class, quality, and blend IAM parameters.  Of the other 5 that were missed, one was excluded because it is blended (despite being classified as a single point source in the SHS), and two did not satisfy the $R-\mathrm{H}\alpha$ vs $R-I$ cut-off criterion.  The remaining two CVs that were not recovered had colours above, but very close to the selection cut.

\begin{figure*}
 \includegraphics[width=178mm]{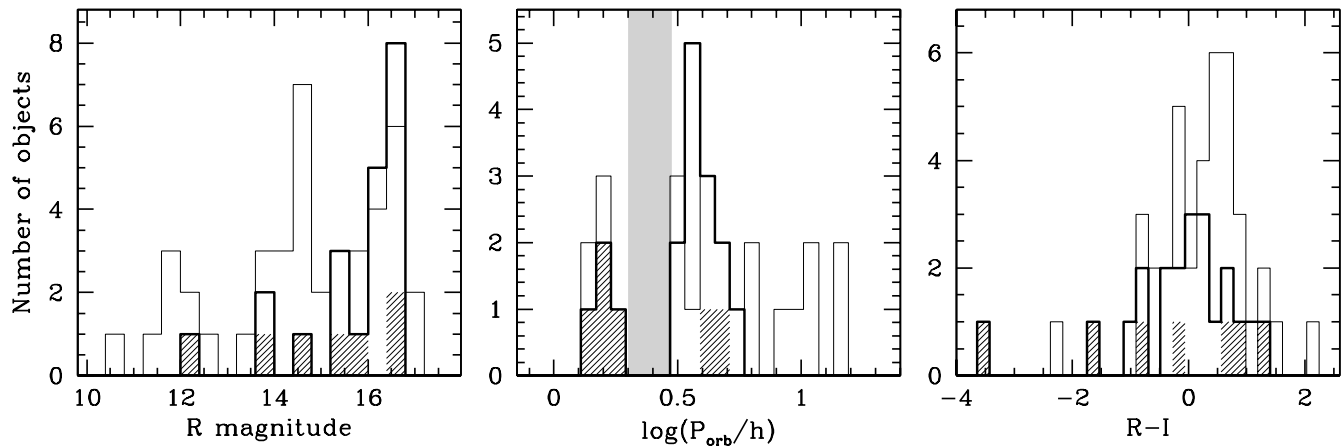}
 \caption{The $R$, $P_{orb}$, and $R-I$ distributions of CVs with $R<17$ in the area covered by our follow-up.  The bold histograms show all systems in the selected sample.  Previously known systems are plotted as fine histograms, and previously known systems recovered by our selection are shown as cross hatched histograms.  The period gap is indicated in the middle panel by light grey shading of the period range 2 to 3~h.  One known CV in our survey area, the nova V1017 Sgr, with $P_{orb}=5.714\,\mathrm{d}$, is not shown on the scale of the $P_{orb}$ histograms. 
}
 \label{fig:per_mag_ri_hist}
\end{figure*}

With 7 recovered systems, and 14 new CVs, we know that our completeness is no better than $(14+7)/(14+42)$, or 38\%.  Since we observed 460 out of 507 targets, the efficiency of selecting previously known CVs is $(7/42)(460/507)$.  I.e., we expect to select roughly 15\% of all CVs at $R<17$, if known CVs occupy the same general part of the $R-\mathrm{H}\alpha$ vs $R-I$ plane as the intrinsic CV population.  Fig.~\ref{fig:per_mag_ri_hist} shows the $R$, $P_{orb}$, and $R-I$ distributions of CVs with $R<17$ in the survey area considered, highlighting the previously known CVs chosen as targets by the selection.

Although the CV recovery rate is low, a more important concern is magnitude- or period-dependent bias in the recovery rate (as that would almost certainly imply the same bias for the overall CV sample).  It is important in this regard that a large fraction of CVs known to have been missed by our selection (30 out of 35) were excluded by the restrictions on IAM parameters, which are expected to be mostly insensitive to colour and apparent magnitude.

The magnitude distributions of both the selected CV sample and the recovered CVs appear to indicate that, up to $R=17$, our survey is not  affected by an apparent magnitude bias.  This is illustrated both by the steep rise towards fainter magnitudes in the $R$ distribution (left hand panel of Fig.~\ref{fig:per_mag_ri_hist}), and by the fact that we do not recover relatively more bright CVs (the same fraction of previously known CVs are recovered in the magnitude bins $R < 14$, $14 \le R < 16$, and $16 \le R < 17$)\footnote{Note that the magnitude distribution of the sample of previously known CVs indicate that it is not approximately complete to $R=17$.  This is not surprising; see, e.g., \cite{PretoriusKniggeKolb07}.}.  The middle panel of Fig.~\ref{fig:per_mag_ri_hist} also shows clearly that a larger fraction of known CVs are recovered at short periods than at long periods.  It is therefore tempting to conclude that, as expected from the fact that short-period CVs have stronger emission lines, any period bias in the selection favours the inclusion of short-period CVs.

The right hand panel of Fig.~\ref{fig:per_mag_ri_hist} shows the distribution of CVs in $R-I$.  The distribution of the sample of previously known CVs has a standard deviation of 1.0.  This large spread in $R-I$ is not physical---it results from poor photometry and variability between the $R$ and $I$ epochs.  The CVs included in our sample are on average bluer than the previously known CVs.  This bias in $R-I$ is not as strong as it appears at first sight.  We will not consider CVs at very long periods, because they are likely to contain evolved secondary stars.  Seven systems in the sample of previously known CVs have periods greater than 8~h, and 5 of these have $R-I$ colours between 0.36 and 0.70, i.e. in the two maximum bins of the $R-I$ distribution.  With these 7 objects excluded, the probability that the $R-I$ distributions of the selected and previously known CVs are drawn from the same population is 0.54, according to a Kolmogorov-Smirnov test.  Therefore, we have no good evidence for a blue bias in $R-I$ from this comparison.  However, the samples used for this comparison are both quite small, and the positive slope of the stellar locus in the $R-\mathrm{H}\alpha$ vs $R-I$ plane, which causes the selection to be more sensitive to objects with relatively small $\mathrm{EW}(\mathrm{H}\alpha)$ at bluer $R-I$, implies that such a bias may be expected.  The implications that this may have for $P_{orb}$- or $R$-dependent biases will be further discussed in Section~\ref{sec:bias}.

\section{The CV sample}
The selection identified a total of 21 CVs; two thirds of this sample consists of newly discovered systems.  Some properties of the CVs comprising the sample are given in Table~\ref{tab:periods}.  The lower limits on the distances ($d_l$) of the previously known CVs were estimated in the same way as described for the new CVs in Paper~I.  The last column of Table~\ref{tab:periods} gives references for the sub-types and orbital periods, where available.

\begin{table}
 \centering
 \begin{minipage}{84mm}
 \caption{Sub-types, orbital periods, $R$-band magnitudes, and lower limits on distances of the CVs included in the survey.  Entries for the systems that will not be included in the comparison to theory are in italics.}
 \label{tab:periods}
 \begin{tabular}{@{}llllll@{}}
 \hline
Object          & Type  & $P_{orb}/\mathrm{h}$& $R$ & $d_l/\mathrm{pc}$ & References\\
 \hline
V1040 Cen       & SU    & 1.45    & 13.9 & 40   & 1  \\
DV Sco          & SU    & 1.65:   & 16.7 & 270  & 2,3\\
\emph{VV Pup}   &\emph{polar}&\emph{1.67392}&\emph{14.6}&\emph{140}&\emph{4,5}\\
CU Vel          & SU    & 1.88:   & 15.8 & 120  & 6  \\
H$\alpha$092134 &       & 3.043   & 16.8 & 530  & 7  \\ 
H$\alpha$073418 & SW:   & 3.18542 & 16.7 & 320  & 7  \\
H$\alpha$074655 &       & 3.3984  & 14.0 & 180  & 7  \\
H$\alpha$102442 &       & 3.673   & 16.4 & 580  & 7  \\ 
H$\alpha$112921 & NL:   & 3.6851: & 15.5 & 510  & 7  \\
H$\alpha$103135 & NL:   & 3.76:   & 16.3 & 630  & 7  \\
H$\alpha$103959 & NL:   & 3.785:  & 16.4 & 670  & 7  \\
H$\alpha$130559 &       & 3.93:   & 16.5 & 500  & 7  \\
H$\alpha$092751 & NL:   & 4.1     & 16.2 & 770  & 7  \\
\emph{CW Mon}   &\emph{DN,IP:}&\emph{4.238}&\emph{15.4}&\emph{210}&\emph{8}\\
H$\alpha$094409 & SW:   & 4.506:  & 16.2 & 520  & 7  \\ 
\emph{HZ Pup}   &\emph{CN,IP}&\emph{5.11}&\emph{16.5}&\emph{1010}&\emph{9}\\
\emph{H$\alpha$074208}& &\emph{5.706}&\emph{15.3}&\emph{490}&\emph{7}  \\ 
H$\alpha$075648 & SW:   & ---     & 16.0 & ---  & 7  \\ 
H$\alpha$163447 & NL:   & ---     & 16.7 & ---  & 7  \\ 
H$\alpha$190039 & NL:   & ---     & 16.8 & ---  & 7  \\ 
V383 Vel        & DN    & ---     & 12.3 & ---  &10  \\
 \hline
 \end{tabular}
References: (1) \cite{Patterson03}; (2) vsnet-alert 8325; (3) Daisaku Nogami, private communication; (4) \cite{Herbig60}; (5) \cite{SchneiderYoung80}; (6) \cite{MennickentDiaz96}; (7) Paper~I; (8) \cite{Kato03}; (9) \cite{AbbottShafter97}; (10) \cite{Williams00}. \\
Notes: The types are dwarf nova (DN); nova like variable (NL); SU UMa star (SU); SW Sex star (SW); intermediate polar (IP); classical nova (CN).  Uncertain values or classifications are denoted by `:'.  H$\alpha$190039 is classified as a CV based only on a single, low quality spectrum. \hfil
\end{minipage}
\end{table}

Although we are not able to provide firm classifications for any of the newly discovered CVs, it is likely that the majority of them are NLs (see Paper~I).  There is no good evidence that any of the new systems are magnetic; however, they have not yet been thoroughly studied.

V1040 Cen, CU Vel, and DV Sco are SU UMa stars.  The orbital period listed for DV Sco is uncertain because it was estimated from the superhump period (the superhump period is 1.71~h, but even this is uncertain, because the superhump photometry available for this system is aliased).  V1040 Cen and CU Vel both have normal fractional superhump period excess, $\epsilon=(P_{sh}-P_{orb})/P_{orb}$, and therefore are not expected to be period bouncers \citep{Patterson03}.  V1040 Cen reaches $V=12.5$ at maximum, and is likely to be very nearby.  The second bright DN in the sample, V383 Vel, was in outburst during the SHS observations.  It is a poorly studied system; the orbital period is not known.

The sample includes two magnetic CVs, namely VV Pup and HZ Pup.  HZ Pup is a classical nova (nova Pup 1963), as well as an IP.  VV Pup is a polar.

\cite{Kato03} report a 37-min modulation in outburst photometry of the DN CW Mon, and suggest that the system might be an IP.  However, the high coherence associated with the spin cycle of a white dwarf has not yet been demonstrated for this signal, and it is detected only near maximum of outburst.  \cite{Warner04} classifies the 37-min modulation as a quasi-period oscillation---a phenomenon commonly seen in high-$\dot{M}$, non-magnetic CVs.  Furthermore, CW Mon does not have particularly strong He\,{\scriptsize II}\,$\lambda$4686 emission \citep{Szkody87}.  Note, however, that the IP HT Cam has a spectral appearance very similar to CW Mon, and an increased white dwarf spin pulse amplitude during outburst \citep{KempPattersonThorstensen02}.  

Magnetic systems are not specifically dealt with in the population model we will use in the next section, and may evolve differently from non-magnetic CVs (e.g. \citealt{WuWickramasinghe93}; \citealt{LiWuWickramasinghe94}; \citealt{WebbinkWickramasinghe02}).  We will therefore exclude the magnetic CVs HZ Pup and VV Pup from the observed sample.  CW Mon is also excluded from the sample, despite the evidence for its magnetic nature not being compelling, because (in view of it being a long-period system) excluding it is a conservative choice.  The conclusion regarding the ratio of long- to short-period CVs that will be derived in Section~\ref{sec:comparison} would be even stronger had we included it in the sample.

We also exclude systems with $P_{orb} > 5\,\mathrm{h}$ from both the model (see Section~\ref{sec:modelling}) and observed populations, since  CVs with evolved secondaries probably dominate the population at these periods (\citealt{BeuermannBaraffeKolb98}; \citealt{BaraffeKolb00}; \citealt{PodsiadlowskiHanRappaport03}).  

Therefore VV Pup, HZ Pup, CW Mon, and H$\alpha$074208 are not considered further.  The entries for these 4 systems are in italics in Table~\ref{tab:periods}.  The rest of the sample (17 CVs, of which at least 10 are long-period systems) likely consists of non-magnetic, unevolved CVs.  There is no indication that any of these CVs are period bouncers.

\section{Modelling selection effects}
\label{sec:modelling}
We use a Monte Carlo technique to quantify the effects of the most important selection biases acting on our sample.  A random sample of CVs is drawn from a predicted intrinsic CV population, and distributed in a model Galaxy.  By modelling the spectral energy distribution (SED) and outburst properties of each CV, we predict the observational appearance of this intrinsic population in a survey with our particular selection criteria.  We use the computational method of \cite{PretoriusKniggeKolb07}, specifically, their model A1, with only a minor change to the SED model.

The Monte Carlo method is described in Section~\ref{sec:mc} below.  It does not account for all the selection biases present in our sample.  In Section~\ref{sec:bias}, we discuss the assumptions that were made to deal with the remaining biases.

\subsection{The Monte Carlo code}
\label{sec:mc}
The Monte Carlo calculation uses as input a probability distribution function (PDF) for the present-day population of CVs over the parameters $P_{orb}$, $\langle \dot{M} \rangle$ (the secular average $\dot{M}$), $M_1$, and $M_2$.  This PDF results from the population synthesis model pm5 of \cite{Kolb93}.  It relies on the predicted CV birth rate of \cite{deKool92} and the magnetic braking prescription of \cite{VerbuntZwaan81}.

The Galaxy is treated as an axisymmetric disc with a Galactic Centre distance of $7\,620\,\mathrm{pc}$ \citep{EisenhauerGenzelAlexander05}, a radial scale length of $3\,000$~pc, and vertical scaleheights of 120, 260, and 450~pc for long-period CVs, pre-period minimum short-period CVs, and period bouncers, respectively.  The assumption of different scaleheights for these three sub-populations of CVs is motivated by their different typical ages, and are chosen to be representative of stellar populations with ages of $\simeq 10^{8.4}$~y, $\simeq 10^{9.2}$~y, and $\simeq 10^{9.6}$~y, respectively \citep{RobinCreze86}; see \cite{PretoriusKniggeKolb07} for a more complete discussion, and for an indication of the sensitivity of the period distribution of magnitude-limited samples to this assumption.

We consider flux from the accretion disc, bright spot, white dwarf, and secondary star in the model of the overall SED of each CV in our simulation, with all except the secondary treated in the same way as in \cite{PretoriusKniggeKolb07}.  The accretion disc contribution is obtained from the model of \cite{Tylenda81}, the bright spot is modelled as a black body, and the white dwarf flux is found from the tabulation of \cite{BergeronWesemaelBeauchamp95}, using the effective temperature predicted by \cite{TownsleyBildsten03}.  The flux contribution of the secondary star is obtained from the semi-empirical donor sequence of \cite{Knigge06}\footnote{We interpolated the high-resolution version of the donor sequence onto the period grid of the model CV population, where possible.  The upper edge of the model period gap is slightly above the observed location.  The model period gap is also narrower than observed, so that we had to extrapolate the empirical sequence at the lower edge of the gap.}.

The only type of variability included in the model is DN outbursts.  The modelling of outbursts relies on the well known empirical relation between $P_{orb}$ and the absolute magnitude of DNe at maximum (e.g. \citealt{brian87}); see \cite{PretoriusKniggeKolb07} for a more complete discussion.  The probability of finding a DN in outburst in a single-epoch survey is then given by the outburst duty cycle ($C$).

The $R$ and $I$ magnitudes produced by our SED model are on the Kron-Cousins $RI$ system, and were transformed to the photographic system of the survey using \cite{MorganParker05} and \cite{Bessell86}.  The differences between the photometric systems are small.

\begin{figure*}
 \includegraphics[width=178mm]{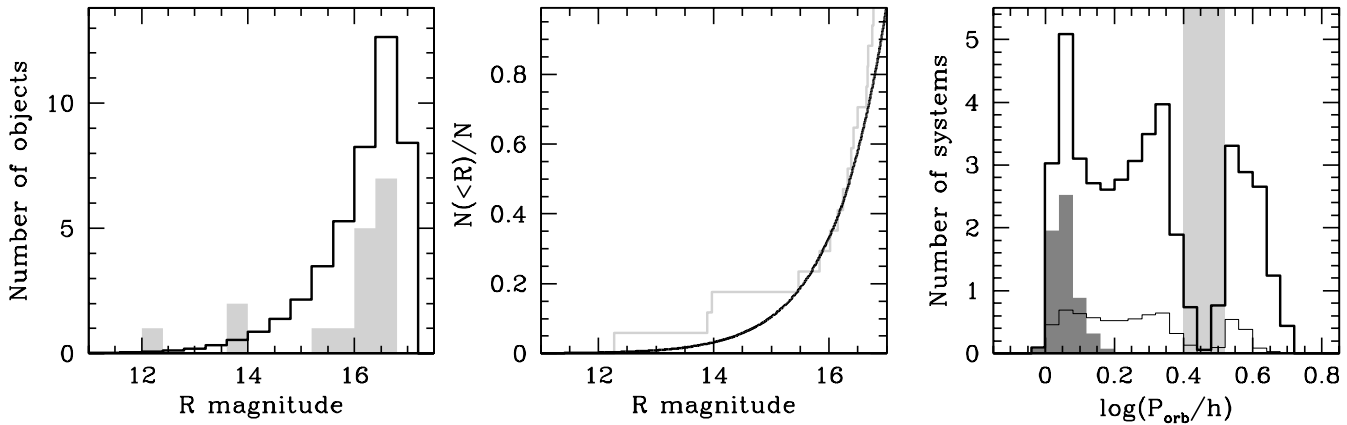}
 \caption{The differential and cumulative magnitude distributions, as well as the differential orbital period distribution, predicted by the simulation.  The bold histograms in all three panels are the distributions of all systems predicted to be detected, with only the flux limit and $|b|$ constraint imposed.  The $R$ and cumulative $R$ distributions of the 17 observed systems are also shown, in light grey, in the left hand and middle panels.  In the right hand panel, the contribution of period bouncers to the orbital period distribution is shaded in dark grey, and the fine histogram shows DNe that are found in outburst in a single epoch.  The model $R$ and $P_{orb}$ distributions are scaled as described in Section~\ref{sec:rho}.}
 \label{fig:model_obs_per_mag}
\end{figure*}

We then select model CVs with $R<17$ and $|b|<10\,\mathrm{deg}$.  Fig.~\ref{fig:model_obs_per_mag} shows the differential and cumulative apparent magnitude distribution, as well as the $P_{orb}$ distribution, of the CV population predicted to be observed in our survey, if we neglect other selection cuts.  The grey histogram in the left hand panel is the $R$ distribution of the 17 observed CVs.  The middle panel shows the cumulative $R$ histograms of the model (black) and real samples (grey).  A Kolmogorov-Smirnov test gives the probability of the model and observed $R$ distributions being drawn from the same parent population as 0.35, implying that the magnitude distributions are consistent.  In the right hand panel, the bold histogram is the $P_{orb}$ distribution of all systems detected; the contribution of period bouncers is shaded in dark grey, and the fine black histogram shows the distribution of outbursting DNe expected in a single-epoch survey.  The model histograms are scaled as described in Section \ref{sec:rho} below.

\subsection{Additional biases}
\label{sec:bias}
The above modelling accounts for the effects of the magnitude limit, and the restriction of the survey to the Galactic plane.  With only those selection effects included, the predicted period distribution is inconsistent with the observations (relatively too many short-period CVs are predicted).  However, we need to determine whether the remaining selection effects present in our sample could reconcile the predicted and observed distributions.  We therefore make conservative assumptions about the influence of these biases (i.e. assumptions that decrease the predicted ratio of short- to long-period CVs).

\subsubsection{The $\mathrm{EW}(\mathrm{H}\alpha)$-dependent selection cut}
The dependence of selection probability on $\mathrm{EW}(\mathrm{H}\alpha)$ is not included in our model.  This is because the disc model predicts EWs that are clearly too small (as demonstrated for H$\beta$ by \citealt{Patterson84}; this is also true for H$\alpha$).  The alternative of using an empirical relation is not available, because the dependence of $\mathrm{EW}(\mathrm{H}\alpha)$ on, e.g., $M_V$ is not known.  One may expect that including this selection cut in the model would increase the predicted ratio of short- to long-period CVs, and therefore that not accounting for it is a conservative assumption.

Whether an object is selected as a target depends on its $R-\mathrm{H}\alpha$ and $R-I$ colours.  The $R-I$ dependence arises from the non-zero slope of the stellar locus  in the $R-\mathrm{H}\alpha$ vs $R-I$ plane.  It is also likely that the selection probability depends on $R$ magnitude, since the photometric uncertainty increases at fainter magnitudes.  While our CV sample does not at first sight appear to be increasingly incomplete at fainter $R$, the sample of all emission line objects from our follow-up certainly is magnitude biased, with an $R$ distribution that flattens off at around $R=16$.  The reason that the CV sample is not affected by the same magnitude bias as the overall emission line sample is not because CVs have systematically stronger emission lines than other types of emission line objects selected for follow-up, but because the CVs are on average bluer in $R-I$.  Our selection is sensitive to faint objects with small EWs only if those objects are sufficiently blue in $R-I$.  Redder faint objects must have stronger lines in order to be selected. 

To the extent that data can guide us, the resulting blue bias does not matter---intrinsically fainter, redder CVs have stronger lines.  This is well established for normal short-period CVs, but observational information on period bouncers is scarce.  There are now two confirmed period bouncers, SDSS J103533.03+055158.4, and SDSS J150722.30+523039.8 (\citealt{LittlefairDhillonMarsh06}; \citealt{LittlefairDhillonMarsh07}; 
%\citealt{PattersonThorstensenKnigge08}
Patterson, Thorstensen \& Knigge (in preparation)) and they have $EW(\mathrm{H}\alpha)=44\,\mathrm{\AA}$ and $78\,\mathrm{\AA}$, and Kron-Cousins $R-I$ of $0.03$ and $0.04$, respectively (\citealt{sdsscvs4}; \citealt{sdsscvs5}; colour transformations from \citealt{Jester05}).  Our selection is easily capable of identifying objects with these EWs and colours near $R=17$.  

In the absence of an undetected population of faint, red CVs with weak emission lines, our EW cut thus preferentially excludes intrinsically bright (i.e. long-period) CVs.  

\subsubsection{Variability}
DNe in outburst probably have smaller EW(H$\alpha$) as a rule (this is certainly true for H$\beta$; \citealt{Patterson84}).  This means that an outbursting DN is less likely to be identified in an H$\alpha$ survey than a quiescent system.  The problem here is slightly more complicated than whether a system has sufficiently large EW(H$\alpha$) at a single epoch, since observations in the three wavebands used in the selection were not taken simultaneously.  Variability between the epochs of the $R$, $I$, and H$\alpha$ observations can lead both to an object with H$\alpha$ emission not being selected, and to an object without H$\alpha$ emission being included.  The most common large amplitude variability displayed by CVs is DN outbursts, and we will consider only this type of variability.  Out of the 175 fields we used in the selection, 72 were observed in $R$ and H$\alpha$ on the same night; all $I$-band data were taken roughly 20 years earlier.

A system that is bright in both $R$ and H$\alpha$ but faint in $I$ will probably be selected, because it will be very blue in $R-I$ (V383 Vel is an example---it was observed in outburst in $R$ and H$\alpha$ on the same night, but was faint during the $I$ epoch).  CVs are likely to be missed by the selection because of variability if they are bright in $R$ and faint in H$\alpha$, or bright only in $I$.  We assume that the probability of a long-period system being excluded for reasons of variability is 0, and that the probability of a short-period system being excluded for being bright in $R$, is $C$.  Then, to allow for the possibility of a DN being bright in $I$, we do the same with the remaining short-period DNe (i.e. we exclude a fraction of systems equal to the particular outburst duty cycle).

\section{Comparison to theory}
\label{sec:comparison}

\subsection{Relative numbers of short- and long-period CVs}
After applying the cuts involving $R$, $|b|$, and variability described above, the sample is predicted to contain 13\% period bouncers (i.e. about 2 systems in our sample of 17), whereas none are detected.  This is not a serious discrepancy.  The predicted fraction of long-period CVs is 30\%.  If we assume that our combined EW and $R-I$ cut makes it impossible to detect period bouncers, then 34\% of the sample is predicted to be long-period systems.  Using the binomial distribution, the probability that the observed ratio of at least $10/17$ is consistent with this prediction is 0.031.  The model is therefore inconsistent with observations at more than 2-$\sigma$.  

The inconsistency would be a 3-$\sigma$ result if two more systems in our sample had periods above the period gap.  It is likely that this will prove to be the case.  The spectrum of H$\alpha$190039 is characteristic of an intrinsically bright CV, and there is reason to suspect that H$\alpha$075648 is an SW Sex star.

\subsection{Space density}
\label{sec:rho}
Our sample is not suitable for measuring the CV space density ($\rho$), but we can verify that it is in reasonable agreement with a specific prediction.  \cite{deKool92} and \cite{Kolb93} theoretically predict $\rho \sim 10^{-4}$ (but see also \citealt{Politano96}, for a theoretical value significantly lower), with a hundredth of all CVs above the period gap.  The space density of long-period CVs ($\rho_l$) is then predicted to be $\sim 10^{-6}\,\mathrm{pc^{-3}}$.  

Setting $\rho_l=10^{-6}\,\mathrm{pc^{-3}}$ in the simulation, we predict (with the area coverage and flux limit of our survey) that about 22 long-period CVs should be detected if the sample really is 15\% complete.  This assumes that the only selection cut affecting long-period CVs is the magnitude limit and $|b|$ restriction.  Our sample contains at least 10 long-period CVs, perhaps as many as 14.  The total number of observed long-period CVs is then consistent with the prediction if the true completeness for long-period systems is $\simeq 8$\%.  This is entirely reasonable, but given the uncertain completeness, it is not a very secure result\footnote{The available observational estimates of $\rho_l$ are $\sim 2.5 \times10^{-6}$ and $\simeq 8 \times10^{-6}$ (\citealt{Patterson98}; \citealt{PretoriusNEP}).}.

The predicted $R$ and $P_{orb}$ histograms in Fig.~\ref{fig:model_obs_per_mag} were scaled to reflect the number of CVs that should have been detected in our survey (2\,800~sq.deg., $|b| < 10^\circ$, and $R<17$) if $\rho_l = 10^{-6}\,\mathrm{pc^{-3}}$, and if the survey is 8\% complete, with no dependence of completeness on $P_{orb}$.

\subsection{Evolutionary time-scales}
\label{sec:tau}
It is usually assumed that the theoretical ratio of the number of short- to long-period CVs is larger than observed because theory predicts too large a population of short-period CVs.  Several solutions to this problem have been proposed.  We will consider here the suggestion that there is an angular momentum loss mechanism that increases the absolute value of $\dot{J}$ of short-period CVs above that of the rate resulting from gravitational radiation ($\dot{J}_{GR}$); see e.g. \cite{Patterson98}; \cite{Patterson01}; \cite{KingSchenkerHameury02}; \cite{Patterson03}; \cite{BarkerKolb03}.  Higher $-\dot{J}$ below the period gap decreases the evolutionary time-scale of short-period CVs and therefore reduces the total number of CVs in this evolutionary phase\footnote{\cite{Patterson98} also points out that a population synthesis model using such a higher $-\dot{J}$ could predict the position of the period minimum correctly, and would fit the observed relation between $\epsilon$ and $P_{orb}$ better than models assuming only gravitational radiation.}.

The evolutionary time-scale of a CV is 
$$\tau \simeq \left \langle \frac{J}{-\dot{J}} \right \rangle \simeq \left \langle \frac{M_2}{\dot{M}} \right \rangle$$
where the angle brackets denote the secular average.  Assuming that all CVs form at periods longer than $\simeq 5$~h, and only evolve through the period ranges that we consider, the number of CVs that exist in some particular period interval is proportional to the time CVs spend in that interval.  The ratio of the numbers of long- and short-period CVs is therefore proportional to the ratio of their evolutionary times-scales ($\tau_l$ and $\tau_s$ for long- and short-period CVs, respectively).  

The ratio of long- to short-period CVs observed in a survey for which the most important bias is a flux limit, can be written as
\begin{equation}
\left(\frac{n_l}{n_s}\right)_{obs} = k(\tau_l/\tau_s) \left(\frac{\tau_l}{\tau_s}\right).
\label{eq:rat1}
\end{equation}
Here, $k > 1$ is a monotonically decreasing function of $\tau_l/\tau_s$, since apparent brightness depends on accretion luminosity, which in turn scales inversely with $\tau$. The form of $k$ means that, by comparing the observed and predicted number ratios, we can set a limit on the true ratio of evolutionary time-scales of CVs above and below the period gap. Using equation~\ref{eq:rat1}, we can write this ratio as
\begin{equation}
\frac{\tau_l}{\tau_s} =
\left(\frac{\tau_l}{\tau_s}\right)_{pred}
\left[
\frac{k((\tau_l/\tau_s)_{pred})}{k(\tau_l/\tau_s)}
\right]
\left[
\frac{\left(n_l/n_s\right)_{obs}}{\left(n_l/n_s\right)_{pred}}
\right],
\end{equation}
where $\left(\tau_l/\tau_s\right)_{pred}$ is the predicted ratio of the evolutionary time-scale of long-period CVs to the evolutionary time-scale short-period CVs.  If $\left(n_l/n_s\right)_{obs} > \left(n_l/n_s\right)_{pred}$, as in our case, we must clearly have 
$\tau_l/\tau_s > \left(\tau_l/\tau_s\right)_{pred}$. The 
monotonic nature of $k$ then implies that 
$k((\tau_l/\tau_s)_{pred}) > k(\tau_l/\tau_s)$, so that
\begin{equation}
\frac{\tau_l}{\tau_s} > \left(\frac{\tau_l}{\tau_s}\right)_{pred}
\left[
\frac{\left(n_l/n_s\right)_{obs}}
{\left(n_l/n_s\right)_{pred}}
\right].
\end{equation}
Physically, this inequality holds because any adjustment to the predicted evolutionary time-scales that causes the short-period phase to be shorter relative to the long-period phase will also make short-period CVs brighter relative to long-period systems (and therefore less under-represented in a flux-limited sample\footnote{The EW cut will not select against short-period CVs, unless their luminosity becomes similar to that of long-period CVs---this means an increase by $\ga10$.}). The impact of such a time-scale adjustment on the observed number ratio is therefore always smaller than on the intrinsic number ratio.

To be more specific, we can assume $\left( \tau_l \right)_{pred}=\tau_l$.  Then, for this survey, with $\left(n_l/n_s\right)_{pred} \simeq 3/7$ and $\left(n_l/n_s\right)_{obs} \ge 10/7$, the model evolutionary time-scale of short-period CVs must be decreased by $\ga3$ to match the observations.  Or, equivalently, we require $-\dot{J} \ga -3 \dot{J}_{GR}$ for short-period CVs, in order to bring theory and observations into agreement.  This factor of at least 3 would be an under estimate if the EW-based selection cut discriminates strongly against long-period CVs.

\section{Discussion}
Despite the very conservative assumptions outlined in Section~\ref{sec:bias}, the model CV population and the observed sample do not agree.  We have thus confirmed the result of \cite{PretoriusKniggeKolb07} with an independent observational sample.  Standard CV evolution theory predicts too large a ratio of short- to long-period CVs.  This is true even if we assume that period bouncers are undetectable.  

The favoured explanation for the large predicted ratio of short- to long-period CVs is that short-period systems evolve faster than predicted by the disrupted magnetic braking model, so that the predicted population of short-period CVs is too large (e.g.\ \citealt{Patterson98}; \citealt{Patterson01}; \citealt{Patterson03}).  Using the H$\alpha$ sample, we find that the evolutionary time-scale of short-period CVs in our model should be decreased by a factor of at least 3 in order for the model to agree with observations (if the predicted evolutionary time-scale of long-period CVs is correct).

For the Palomar-Green (PG) Survey CV sample (\citealt{GreenSchmidtLiebert86}; \citealt{Ringwald93}), \cite{PretoriusKniggeKolb07} found $\left(n_l/n_s\right)_{pred} \simeq 7/93$, compared to an observed ratio of $\left(n_l/n_s\right)_{obs} \ge 14/13$.  With the same reasoning as in Section~\ref{sec:tau} above, this implies that the evolutionary time-scale of short-period CVs in the model is too long by more than an order of magnitude.  This is surprisingly large, especially when compared to the result of \cite{Patterson01}, who finds that $\dot{J}=3\dot{J}_{GR}$ for short-period CVs provides a satisfactory fit to his data.  This may be a hint that the assumption $\left( \tau_l \right)_{pred}=\tau_l$ is wrong.  The observational constraints on the size of the long-period CV population and $\dot{M}$ above the period gap are not sufficiently precise to answer this question definitively, but it certainly has been suggested that the \cite{VerbuntZwaan81} magnetic braking law used in the model is too strong (e.g. \citealt{IvanovaTaam03}).  Our model may therefore not only predict too many short-period CVs, but also too few long-period CVs.  

In this regard, it is also worth noting that recent estimates of $\dot{M}$ in a small sample of short-period CVs are consistent with the values expected from angular momentum loss through gravitational radiation (\citealt{LittlefairDhillonMarsh06}; Littlefair, et al.\ (in preparation)).  \cite{LittlefairDhillonMarsh07} have also estimated $\dot{M}$ for SDSS J150722.30+523039.8, but they find that this system probably has an unusual evolutionary history (see also Patterson, Thorstensen \& Knigge (in preparation)).

Before an H$\alpha$-selected CV sample can be used to  place tighter constraints on CV evolution, $\mathrm{EW}(\mathrm{H}\alpha)$ must be modelled in terms of the parameters describing CVs in a theoretical population.  It is known that the EWs of disc emission lines increase with decreasing luminosity.  However, CV discs are observed together with a white dwarf primary.  Very faint systems are white dwarf dominated (i.e., the continuum emission from the disc is low relative to that from the white dwarf).  Also, the disc emission lines are superimposed on the absorption lines of the white dwarf.  Therefore, at faint luminosity, the relation between EW and $M_V$ is expected to flatten off (\citealt{Patterson84} plots $\mathrm{EW}(\mathrm{H}\beta)$ vs the absolute magnitude of the disc alone; his relation therefore does not show this effect).  This is observationally clear from the spectra of the faintest known CVs, and has already been remarked on by \cite{AungwerojwitGansickeRodriguez-Gil06}.  Although the decrease in $\mathrm{EW}(\mathrm{H}\alpha)$ is expected to be less severe than that in $\mathrm{EW}(\mathrm{H}\beta)$, this might ultimately place a limit on what can be achieved by an emission line survey.

The comparison of this H$\alpha$-selected CV sample to theory is also complicated by the possibility of variability between non-simultaneous observations in the three wave-bands used for the selection.  Surveys such as IPHAS will overcome this difficulty.  IPHAS will also provide much better photometric accuracy than can be achieved photographically, so that the photometry will allow for the identification of objects with smaller EWs in all parts of the colour-colour plane; any blue bias will therefore be weaker (see \citealt{Drew05}; \citealt{WithamKniggeGansicke06}; \citealt{WithamKniggeAungwerojwit07}; \citealt{WithamCat}).  Furthermore, the emission line samples constructed from IPHAS data will be much deeper.  This is very important, since a selection technique capable of identifying intrinsically faint CVs is of limited value in a survey with a bright flux limit.  The main challenge for IPHAS will be the amount of observational effort needed to finish the identification and follow-up of the CV sample.  The same is true for other large, deep surveys---the completion of the resulting new CV samples are still some time in the future.  

With the smaller CV samples available at the moment, it is already possible to recognize the qualitative changes that are needed to reconcile theory with observations.

\section{Conclusions}
We have compared a homogeneous CV sample, selected for H$\alpha$ emission, to a model CV population based on standard CV evolution theory.  The magnitude limit and Galactic latitude range of the observed sample was modelled in some detail, while conservative assumptions were made to account for the effects of variability and the $\mathrm{EW}(\mathrm{H}\alpha)$-based selection cut.  The model population is inconsistent with the observed sample.  Specifically, the model predicts relatively too many short-period CVs.  This confirms earlier results, based on independent observations.

The reason for the mismatch between the predicted and observed ratio of short- to long-period CVs may be that the theoretical evolutionary time-scale for CVs below the period gap is too long.  A (very simplistic) consideration of the relative numbers of long- and short-period CVs included in the sample indicates that the disrupted magnetic braking model underestimates $-\dot{J}$ of short-period CVs by a factor of at least 3, assuming that the model is correct for long-period CVs.

Although surveys now in progress will in future provide much better observational constraints on CV evolution theory than can be derived at the moment, it is already clear that the standard magnetic braking model is in need of revision.  Furthermore, it seems that the correct approach to take is to investigate angular momentum loss rates in excess of the gravitational radiation rate in CVs below the period gap.

\section*{Acknowledgements}
MLP acknowledges financial support from the South African National Research Foundation and the University of Southampton.  We thank Ulrich Kolb for providing the model CV population used here, Romuald Tylenda for the use of his accretion disc model, and Daisaku Nogami for making the results of observations of DV Sco available to us.  This paper was improved by the comments of the referee, Stuart Littlefair.

\bsp

\label{lastpage}

\end{document}